\documentclass[%
pre,preprint,
superscriptaddress,
tightenlines,
showpacs,showkeys,
a4paper,12pt
]{revtex4}
\usepackage{dcolumn}
\usepackage{amssymb,amsmath,array}
\usepackage{graphicx}
\usepackage{subfigure}

\renewcommand{\thesubfigure}{(\alph{subfigure})}

\makeatletter
\renewcommand{\@thesubfigure}{\thesubfigure\space}
 
\newcommand{\sca}[2]{\bigl({#1}\cdot{#2}\bigr)}
\newcommand{\avr}[1]{\left\langle{#1}\right\rangle}

\newcommand{\bs}[1]{\boldsymbol{#1}}
\newcommand{\vc}[1]{\mathbf{#1}}
\newcommand{\uvc}[1]{\mathbf{\hat #1}}

\newcommand{\dd}{\mathrm{d}}
\newcommand{\e}{\mathrm{e}}
\newcommand{\eff}{\mathrm{eff}}

\newcommand{\mrm}[1]{\mathrm{#1}}

\newcommand{\Tr}{\mathop{\rm Tr}\nolimits}

\makeatother

\begin{document}
\DeclareGraphicsExtensions{.jpg,.pdf}

\title{%
Ordering of droplets and
light scattering in polymer dispersed liquid crystal films
}

\author{A.D.~Kiselev}
\email[Email address: ]{kisel@mail.cn.ua}
\affiliation{%
 Chernigov State Technological University,
 Shevchenko Street 95,
 14027 Chernigov, Ukraine
} 

 \author{O.V.~Yaroshchuk}
 \email[Email address: ]{olegyar@iop.kiev.ua}
 \affiliation{%
 Institute of Physics of NASU, prospekt Nauki 46,
 03028 Ky\"{\i}v, Ukraine
}

\author{L. Dolgov}
\affiliation{%
 Institute of Physics of NASU, prospect Nauki 46,
 03028 Ky\"{\i}v, Ukraine
} 

\date{\today}

\begin{abstract}
We study the effects of droplet ordering
in initial optical transmittance through 
polymer dispersed liquid crystal
(PDLC) films
prepared in the presence of an electrical field.
The experimental data are interpreted by using
a theoretical approach to 
light scattering in PDLC films
that explicitly relates optical transmittance  and  the order parameters 
characterizing both the orientational structures
inside bipolar droplets and orientational distribution of the droplets.
The theory relies on the Rayleigh-Gans approximation and 
uses the Percus-Yevick approximation to
take into account the effects due to droplet positional correlations.
\end{abstract}

\pacs{%
61.30.-v, 42.79.Kr, 42.25.Fx, 77.84.Lf, 78.66.Sq
}
\keywords{%
light scattering; polymer dispersed liquid crystal; 
orientational ordering
} 
 
\maketitle

\section{Introduction}
\label{sec:intro}

Polymer dispersed liquid crystal (PDLC) films consist of
randomly distributed
micrometer-sized liquid crystal (LC) droplets embedded in an isotropic
polymer matrix, and have attracted considerable interest
for both technological and for more fundamental 
reasons~\cite{Doan:1990,Drza:1995,Craw:bk:1996,Hig:adv:2000}.

The droplets are typically filled with a nematic
liquid crystal (NLC). Optical characteristics of such
birefringent nematic droplets differ from those of 
the surrounding polymer matrix
and the droplets can be viewed as optically anisotropic inhomogeneities
which scatter light incident upon them.

Optical transmittance of PDLC films is mostly determined by
light scattering properties of NLC droplets that crucially depend on NLC
orientational structure.
This structure  
can be influenced by an external electrical field
which reorients the NLC directors inside the droplets
and thereby light scattering in the film appears to be governed by the field.
This mechanism underlies the mode of operation of PDLC films
that, under certain conditions, can be switched from an opaque to
a clear state by applying an electric voltage. 

\begin{figure*}[!tbh]
\centering
\subfigure[Off state with $Q=0$]{%
\resizebox{80mm}{!}{\includegraphics*{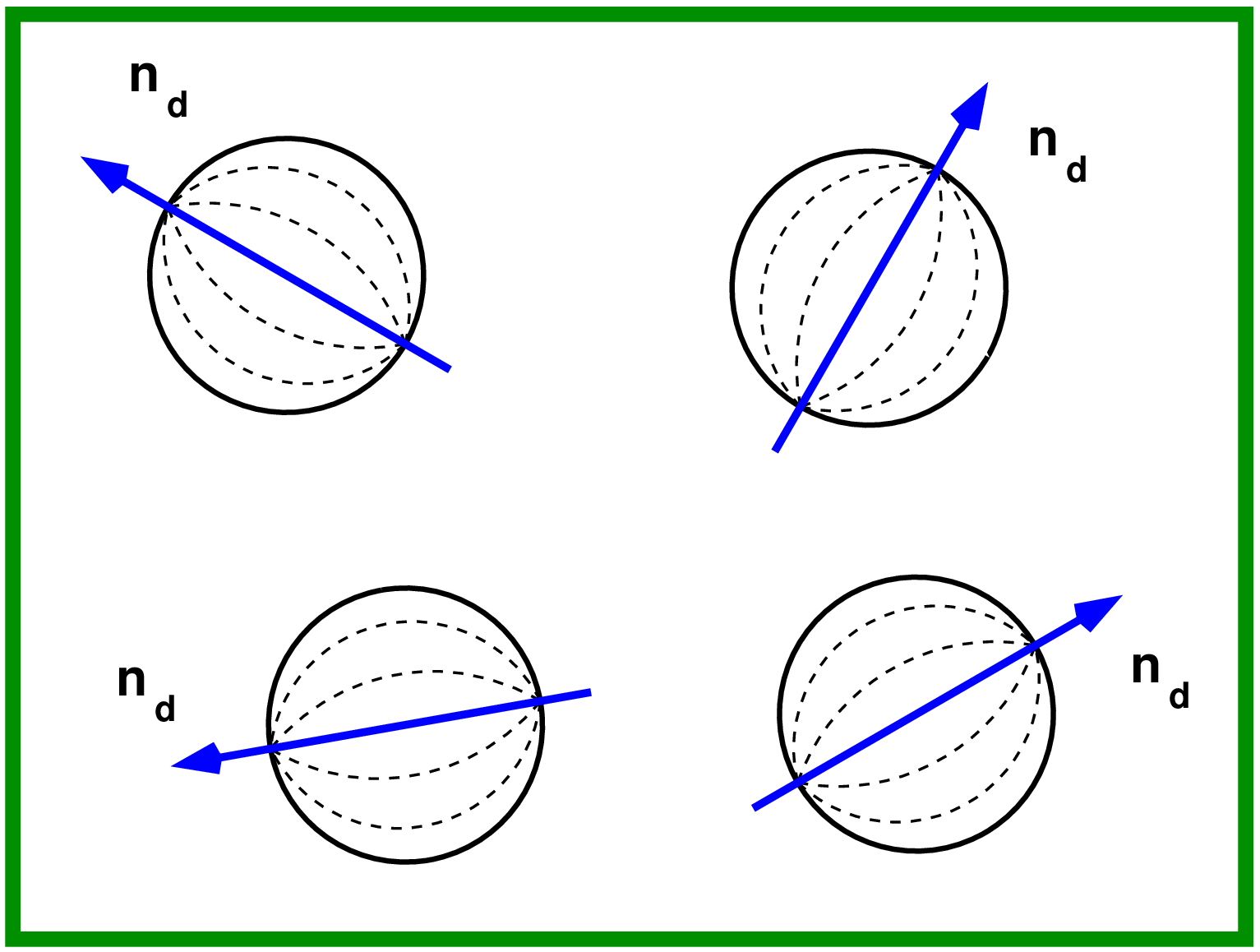}}
\label{fig:disord}}
\subfigure[On state with $Q\approx 1$]{%
\resizebox{80mm}{!}{\includegraphics*{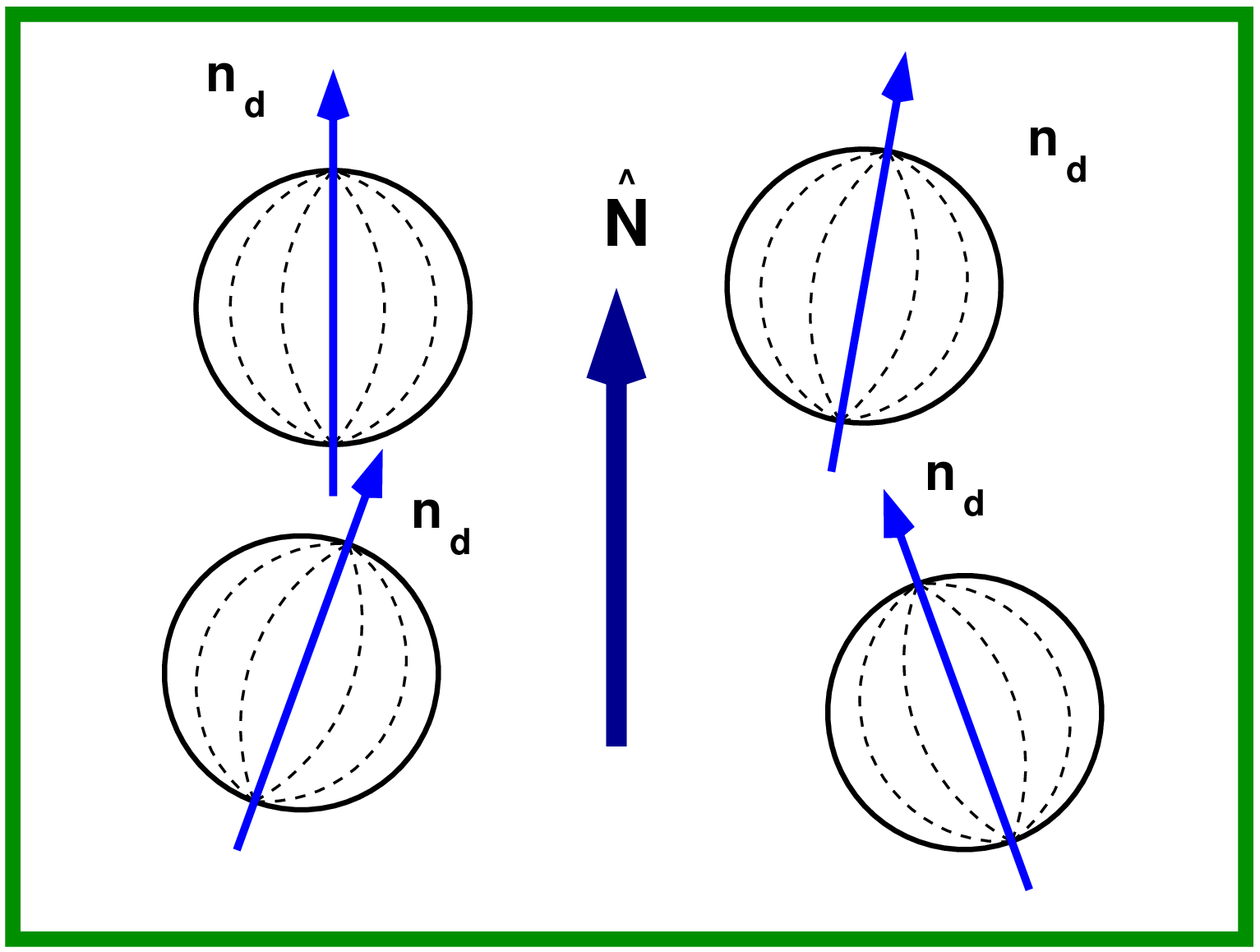}}
\label{fig:ord}}
\caption{%
(a)~Random and (b)~uniaxially 
anisotropic orientational distributions of bipolar droplets.  
}
\label{fig:nd-distr}
\end{figure*}

Schematic representation of the field effect 
on the orientational structure is given in
Fig.~\ref{fig:nd-distr}.
The zero-field case in which 
orientation of NLC is randomly distributed over the droplets
is shown in Fig.~\ref{fig:disord}.
As is illustrated in Fig.~\ref{fig:ord},
in the presence of a field, the NLC directors align along the
prescribed direction $\uvc{N}$. 
When the polymer index of refraction, $n_p$, and 
the ordinary refractive index of NLC, $n_{lc}^{(o)}$,
are matched, the PDLC film in the on state is almost transparent
for waves propagating along the voltage induced anisotropy axis
$\uvc{N}$.

Optical switching contrast, which is the ratio
of the transmittance in the on state and the initial (zero-field)
transmittance, is an important characteristic of the PDLC film.
High switching contrast can be achieved by reducing 
the zero-field transmittance of the PDLC film and,
for this purpose, the film should be prepared so as to
maximize optical contrast between the polymer and NLC.

Typically, the extraordinary refractive index $n_{lc}^{(e)}$ 
is the largest value of the NLC refractive index  
and $n_{lc}^{(o)}\approx n_{p}< n_{lc}^{(e)}$.
So, for light normally incident upon the film,  
in-plane alignment of NLC inside the droplets will enhance
scattering efficiency.
The experimental procedures
for stimulating NLC molecules to be aligned in the plane of the PDLC film
were suggested in Refs.~\cite{Doan:1990,Marg:lc:1985,Wu:phla:1997}.
In these methods
the film of PDLC precursor was subjected to external fields
[mechanical pressure, light, electrical
and magnetic fields] during the phase separation.   

In this paper we present
both experimental and theoretical results for
initial transmittance of PDLC films prepared in the presence of an
electrical field applied across the film of PDLC precursor.
In this case the voltage will align NLC molecules perpendicular to the film
thus reducing the switching contrast.
We are, however, primarily interested in studying
relation between the optical transmittance of a PDLC film and the
order parameters characterizing both the orientational structures
inside the droplets and overall anisotropy of the film.

%% It is also commonly accepted that PDLC films prepared in the absence of
%% external fields are isotropic.
%% This assumption, however, has not been tested experimentally yet. 

To this end, we need to have a theory that
explicitly relates the order parameters and the transmittance
to interpret the experimental data. 
In Refs.~\cite{Vic:mclc:1995,Vic:opt:1996} the
theory by Kelly and Palffy-Muhoray~\cite{Kel:mclc:1994},
that studied the electrical field induced effects in light
scattering by an ensemble of bipolar droplets,
has been used to discuss the order
parameter effects in optical transmittance through a PDLC film.

%% The light scattering problem for an ensemble of nematic droplets was
%% originally considered by \v{Z}umer \textit{et al.}  in
%% Ref.~\cite{Zum:josa:1989}.  Subsequently, Kelly and Palffy-Muhoray
%% theoretically studied the electrical field induced effects in light
%% scattering by a collection of bipolar droplets~\cite{Kel:mclc:1994}.

In the present paper these effects will be studied by using a more
comprehensive approach that takes into account 
interference effects caused by droplet positional correlations. 
As in Ref.~\cite{Cox:1998}, for this purpose,
we shall use the Percus-Yevick approximation.
We also combine the effective medium theory~\cite{Levy:2000}
with the low concentration approximation to account for 
the dependent scattering effects.

The paper is organized as follows.
In Sec.~\ref{sec:experiment}, we give 
necessary details on the experimental
procedure used in this study and describe the morphology
of the PDLC system under consideration. 
We present
the experimentally measured dependencies of
the zero-field transmittance and the transmittance in the saturated on state
on the voltage $U_{uv}$ applied across the film during UV
polymerization.

Our theoretical considerations are presented in
Sec.~\ref{sec:theory}.
We characterize
NLC director structures inside the bipolar droplets
and the orientational distribution of the bipolar axis
by the bipolar order parameter $Q_{\dd}$ and 
the order parameter $Q$.
Then we compute the effective dielectric constant
and use the Rayleigh-Gans approximation to 
evaluate the scattering cross section
of a bipolar droplet embedded in the effective medium.
After averaging the cross section over both positions and orientation
of bipolar droplets we derive the scattering mean free path and
the optical transmittance.
Dependencies of the transmittance on the order parameters
at different droplet sizes are calculated numerically. 

In Sec.~\ref{sec:discussion}, 
we draw together and discuss our results. 
The experimental data are interpreted by using the theoretical results.
The order parameters in the off state are estimated.
The value of the bipolar order parameter $Q_{\dd}$
is found to be about $0.85$.
For samples prepared with no applied voltage, 
the order parameter $Q$ turns out to be very close to zero.
This order parameter rapidly grows and saturates as
the voltage $U_{uv}$ increases.
Finally, we conclude in Sec.~\ref{sec:concl}.

\begin{figure*}[!tbh]
%\vskip5mm
\centering
\resizebox{150mm}{!}{\includegraphics*{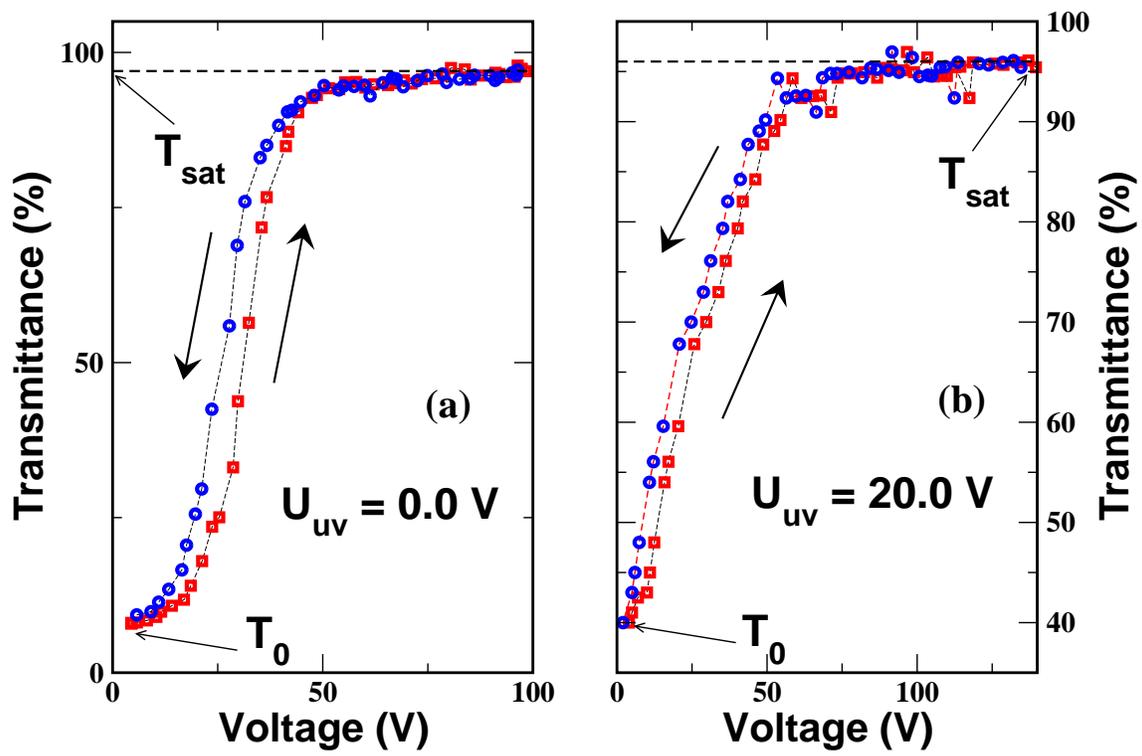}}
\caption{%
$T$-$U$ dependences for two PDLC samples prepared at 
(a)~$U_{uv}=0$~V and (b)~$U_{uv}=20$~V.
Arrows indicate an increase and a decrease of
the applied voltage $U$. 
}
\label{fig:trans-exp-u}
\end{figure*}

\begin{figure*}[!tbh]
%\vskip5mm
\centering
\resizebox{120mm}{!}{\includegraphics*{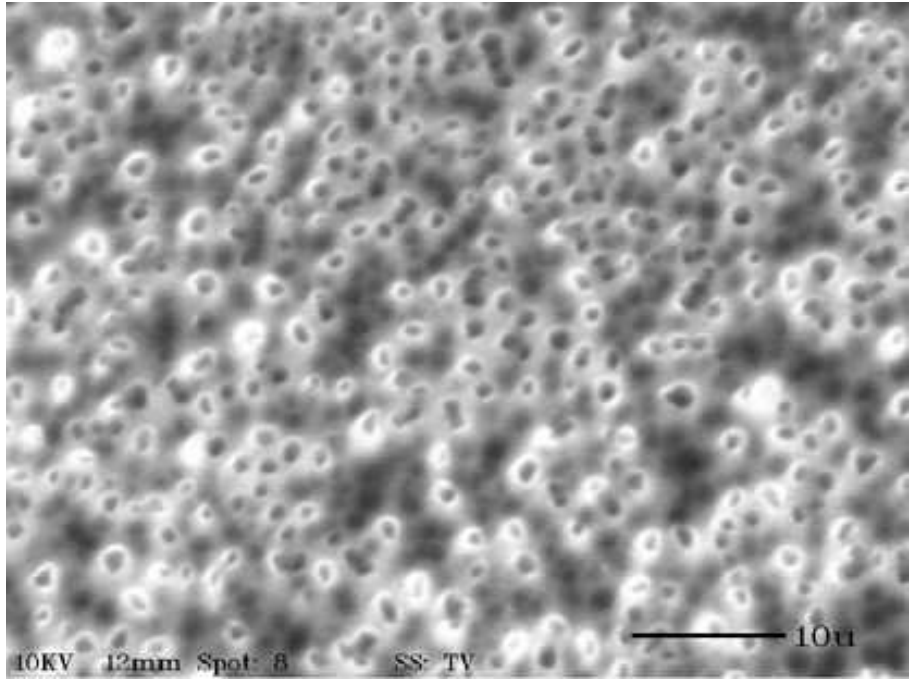}}
\caption{%
SEM image of PDLC layer ($c_p=60$~wt\%, $c_{lc}=40$~wt\%).
}
\label{fig:morph}
\end{figure*}

 \begin{figure*}[!tbh]
% \vskip5mm
 \centering
 \resizebox{150mm}{!}{\includegraphics*{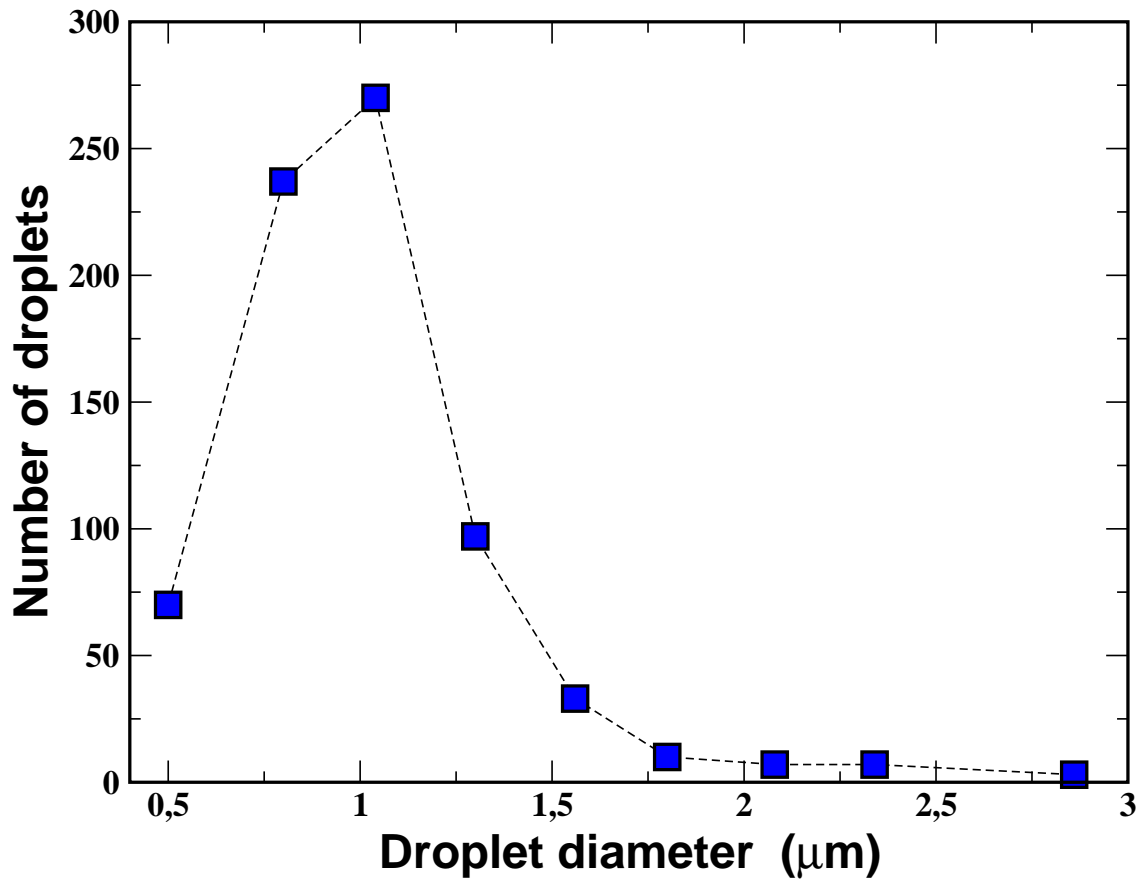}}
 \caption{%
Distribution of droplet sizes in PDLC layer
($c_p=60$~wt\%, $c_{lc}=40$~wt\%).  
 }
 \label{fig:size-numb}
 \end{figure*}

\begin{figure*}[!tbh]
\centering
\subfigure[$U_{uv}=0$~V]{%
\resizebox{80mm}{!}{\includegraphics*{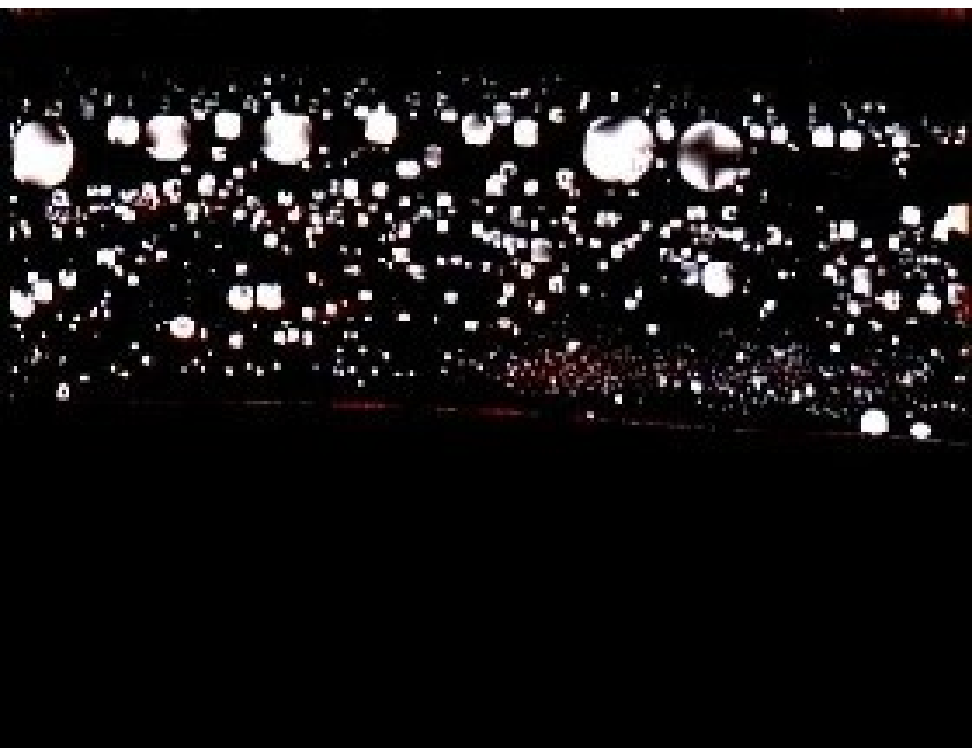}}
\label{fig:Uuv0}}
\subfigure[$U_{uv}=40$~V]{%
\resizebox{80mm}{!}{\includegraphics*{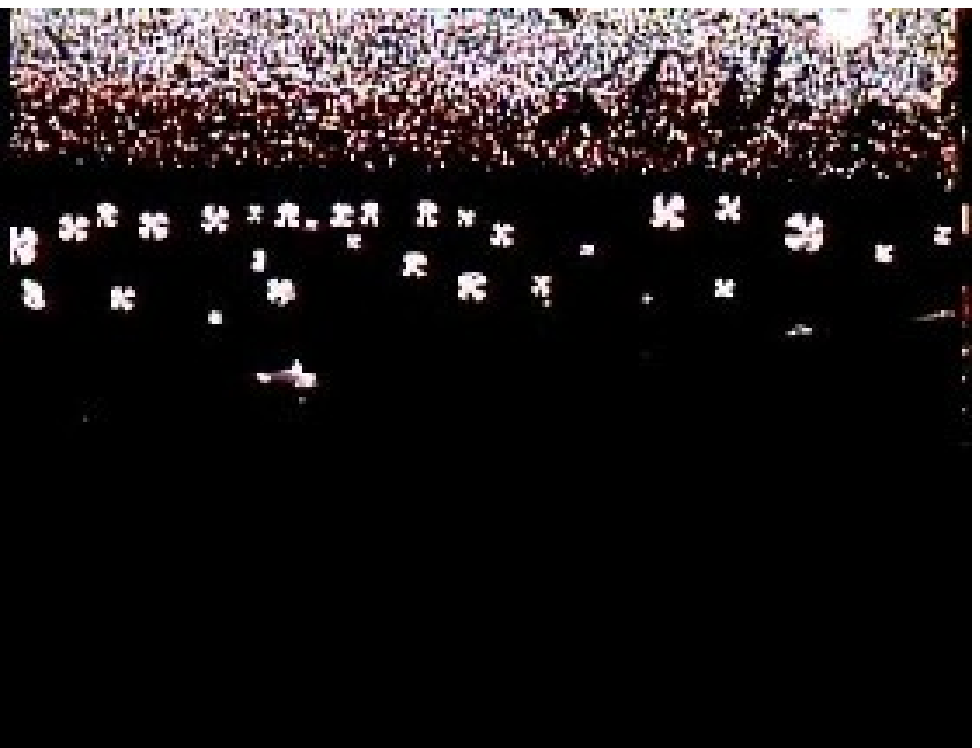}}
\label{fig:Uuv40}}
\caption{%
Periphery part of PDLC layers as viewed in polarizing microscope. 
The layers are preparated at  (a)~$U_{uv}=0$~V and (b)~$U_{uv}=40$~V. 
It is clearly seen that the bipolar droplets 
are orientationally disordered at $U_{uv}=0$~V 
and are oriented along the field at $U_{uv}=40$~V.   
}
\label{fig:periph}
\end{figure*}

\section{Experimental}
\label{sec:experiment}

\subsection{Sample preparation}
\label{subsec:samples}

NLC mixture E7 from Merck and UV
curable adhesive POA65 from Norland Inc.(USA) were used 
to prepare the composite.
PDLC systems based on these  
components have been studied intensively
~\cite{Wu:phla:1997,Bhar1:macr:1999,Bhar2:macr:1999,Bhar3:macr:1999,Dolg:spie:2003}. 

The components were thoroughly mixed. The content
of LC and polymer precursor in the mixture was 
$c_p=60$~wt\% and $c_{lc}=40$~wt\%,
respectively. The mixing was performed under red light so as to avoid early
polymerization of POA65. The prepared mixture was sandwiched between
two glass slabs covered with ITO on the inner side. The cell gap was
maintained by spacer balls 20~$\mu$m in size. 
The substrates were
pressed and glued with a epoxy glue.
 
In order to stimulate
polymerization-induced
phase separation leading to the formation of
PDLC layers 
the samples were irradiated with the light from mercury lamp. 
The intensity and the time of irradiation were 
100 mW/cm$^2$ and 20~min, respectively. 

The samples were subjected to
the electrical field $U_{uv}$ of the frequency 
$f=2$~kHz during UV polymerization.
The voltages $U_{uv}=$0~V, 2~V, 10~V, 20~V, 50~V and 100~V
were applied across the samples to prepare
PDLC layers with varying orientational distribution of droplets and,
as a result, initial transmittance.
The difference in transmittance was visible even to the human eye.

SEM measurements were carried out using device S-2600N from Hitachi.
For performing the  measurements the samples were treated as
follows. One glass 
slab was removed and the remaining part of the sample was immersed into
ethanol of purity 98\% for 24 h. Then the sample was taken out from
ethanol, dried and coated with gold.

\subsection{Optical transmittance}
\label{subsec:exper-set-up}

Optical transmittance $T$ versus applied voltage $U$ curves were
measured at room temperature using the computer conjugated measuring
system previously described in Ref.~\cite{Zakr:mclc:2002}. 
The sample transmittance was defined as the ratio: 
\begin{equation}
  \label{eq:T-exper}
  T=\frac{I_{out}}{I_{in}},
\end{equation}
where $I_{in}$ is the intensity of
the normally incident light with the wavelength $\lambda=635$~nm 
and $I_{out}$ is the intensity of
the light transmitted through the sample. 
Following
usual procedure, the non-scattered light and the light scattered
within the angle of 2 degrees 
were detected with a photodiode. The $T$-$U$ curves
were measured at both increasing and decreasing voltage. 
In this process, the voltage was varied from
0~V to 100~V, whereas the frequency was fixed at 2~kHz. Typical $T$-$U$
curves are shown in Fig.~\ref{fig:trans-exp-u}. 
As is indicated,
the curves provide transmittance in the initial
zero-field state ($T_0$) and in a state of saturation 
reached at sufficiently large voltages ($T_{sat}$). 

\begin{table*}[htbp]
\begin{ruledtabular}
\begin{tabular}{D{.}{.}{-1}D{.}{.}{-1}D{.}{.}{-1}|D{.}{.}{-1}D{.}{.}{-1}|D{.}{.}{-1}D{.}{.}{-1}}
\multicolumn{3}{c|}{Experiment}&
\multicolumn{4}{c}{Theory }\\
\colrule
\multicolumn{1}{c}{$U_{uv}$ (V)}&
\multicolumn{1}{c}{$T_0$ (\%)}&
\multicolumn{1}{c|}{$T_{sat}$ (\%)}&
\multicolumn{2}{c|}{$Q_{\dd}=0.85$}&
\multicolumn{2}{c}{$Q_{\dd}=0.9$}
\\
\cline{4-7}
\multicolumn{1}{c}{expt.}&
\multicolumn{1}{c}{expt.}&
\multicolumn{1}{c|}{expt.}&
\multicolumn{1}{c}{$T_0$ (\%)}&
\multicolumn{1}{c|}{$Q$}&
\multicolumn{1}{c}{$T_0$ (\%)}&
\multicolumn{1}{c}{$Q$}
\\
\colrule
0 & 8 & 97& 9.0 & 0.04 & 7.0 & 0.05
\\
2 & 10 & 97& 10.04 & 0.25 & 10.3 & 0.4
\\
10 & 13 & 90& 12.9 & 0.4 & 13.0 & 0.49
\\
20 & 40 & 95& 39.8 & 0.775 & 40.5 & 0.8
\\
50 & 84 & 99& 83.5 & 0.96 & 83.6 & 0.97
\\
100 & 88 & 99& 87.5 & 0.97 & 89.0 & 0.98
\\
\end{tabular}
\end{ruledtabular}
\caption{%
Zero-field and saturated transmittance, $T_0$ and $T_{sat}$,
measured at different values of the voltage $U_{uv}$ applied during
polymerization 
(the film thickness is $20\,\mu$m and the accuracy of the measurements
is about $\pm 0.7$\%).
The corresponding
theoretical values of $T_0$ and the order parameter $Q$
are given in the last four columns.
}
\label{tab:T-Q-vs-U}
\end{table*}

\subsection{Experimental results}
\label{subsec:exper-res}

A typical SEM image and the corresponding distribution of droplet sizes
are shown in Fig.~\ref{fig:morph} 
and in Fig.~\ref{fig:size-numb}, respectively. It is seen that the
preparation procedure yields the well-known ``Swiss cheese'' PDLC morphology
with a uniform spatial distribution of 
the droplets. We did not observe an
anisotropy of droplet shape in the probes prepared in the presence
of the field.  
The droplet size distribution is rather narrow. The
prevailed droplet diameter is about 1~$\mu$m.

The director configuration inside NLC droplets 
with diameters of about 5~$\mu$m can be identified by observation in
polarizing microscope. Such large droplets often form 
in outlying parts of PDLC layers during phase separation. 
The pictures typical of PDLC 
formed at $U_{uv}=0$~V and $U_{uv}=40$~V are shown 
in Fig.~\ref{fig:Uuv0} and in Fig.~\ref{fig:Uuv40}, respectively. 
It is seen that the droplets are spherically shaped 
regardless of
the field applied during phase separation. 
The bipolar structure inside the droplets 
can also be clearly observed. 
For the sample prepared at 
$U_{uv}=0$, the droplets are randomly oriented, 
whereas the droplets in the sample obtained at 
$U_{uv}=40$~V are aligned along the direction of 
the applied field.

For the concentrations  
of the components used in our samples (60~wt\% of polymer and
40~wt\% of NLC), the concentration of NLC 
droplets is not large. It is
confirmed by weak scattering of light in these composites.  

For PDLC based on our components,
the fractions of NLC confined in droplets 
and of NLC dissolved in the polymer matrix
were obtained from IR measurements in Ref.~\cite{Bhar2:macr:1999}. 
According to these results, in our case
only about 30~\% of NLC phase separates. 
Assuming that the density of
polymer is 1.5 g/cm$^3$ and the density of E7 is 1.05 g/cm$^3$, 
we can estimate the volume fraction of droplets, $\eta$,
to be about 23.5~vol\%. 

The obtained data are summarized in Table~\ref{tab:T-Q-vs-U}, 
where the values of
the transmittance in the off state $T_0$
and the transmittance in the saturated state $T_{sat}$ 
measured for different values of
the voltage $U_{uv}$ applied
during phase separation. 
It can be concluded that 
the transmittance in the saturated state
$T_{sat}$ is almost insensitive to the voltage 
$U_{uv}$ applied during phase separation 
as opposed to 
the initial transmittance $T_0$ which
rapidly grows with $U_{uv}$ and eventually saturates.

\section{Theory}
\label{sec:theory}

\subsection{Model and order parameters}
\label{subsec:model}

In this section our task is to study the problem of
light scattering by an ensemble of partially ordered
spherical NLC droplets dispersed 
in an optically isotropic polymer
matrix with the dielectric constant $\epsilon_p$
and the refractive index $n_p=\sqrt{\epsilon_p}$.
The ensemble of such droplets can be regarded as a
reasonably simplified model of a PDLC system.

Specifically, we consider 
the case in which NLC is tangentially ordered at
the droplet surfaces and the NLC director field
$\uvc{n}_{\mrm{LC}}$ inside the droplet 
forms the bipolar orientational structure.
The symmetry of
this structure is cylindrical and 
there are boojum surface singularities at the droplet poles
defined by the symmetry (bipolar) axis 
of the bipolar configuration $\uvc{n}_\dd$.

The bipolar droplets are optically anisotropic
with the dielectric tensor
\begin{equation}
  \label{eq:eps-lc}
  \bs{\epsilon}_{\mrm{LC}}=
\epsilon_{\perp}^{(lc)}\,\vc{I}+
\epsilon_a^{(lc)}\,
\uvc{n}_{\mrm{LC}}\otimes\uvc{n}_{\mrm{LC}},
\end{equation}
where
$\epsilon_a^{(lc)}=\epsilon_{\parallel}^{(lc)}-\epsilon_{\perp}^{(lc)}$
and $\vc{I}$ is the identity matrix.
Thus, the refractive indices of ordinary and extraordinary waves
are:
$n_o^{(lc)}=\sqrt{\epsilon_{\perp}^{(lc)}}$
and
$n_e^{(lc)}=\sqrt{\epsilon_{\parallel}^{(lc)}}$, respectively.

We shall also need to introduce the dielectric tensor
averaged over the director distribution inside a droplet
$\avr{\bs{\epsilon}_{\mrm{LC}}}_{lc}\equiv\bs{\epsilon}_{\dd}$.
Owing to the cylindrical symmetry
of the orientational structure, 
each bipolar droplet can be characterized by the nematic-like 
tensorial order parameter
\begin{equation}
  \label{eq:ord-drp}
  \avr{(3 \uvc{n}_{\mrm{LC}}\otimes\uvc{n}_{\mrm{LC}}-\vc{I})/2}_{lc}
=Q_{\dd}(3 \uvc{n}_{\dd}\otimes\uvc{n}_{\dd}-\vc{I})/2,
\end{equation}
so that the average dielectric tensor of a bipolar droplet
$\bs{\epsilon}_{\dd}$ is
\begin{equation}
  \label{eq:eps-drp}
  \bs{\epsilon}_{\dd}=
\avr{\bs{\epsilon}_{\mrm{LC}}}_{lc}=
\epsilon_{\perp}\,\vc{I}+
\epsilon_a\,\uvc{n}_{\dd}\otimes
\uvc{n}_{\dd},
\end{equation}
where
$
\epsilon_{\perp}=\epsilon_{\perp}^{(lc)}+
\epsilon_{a}^{(lc)} (1-Q_{\dd})/3
$
and $\epsilon_{a}\equiv\epsilon_{\parallel}-\epsilon_{\perp}=
\epsilon_{a}^{(lc)} Q_{\dd}$.

By analogy to the bipolar axis $\uvc{n}_d$
which defines the droplet orientation,
the scalar order parameter $Q_{\dd}$ will be referred to as the
bipolar order parameter.
This parameter characterizes the degree of
droplet anisotropy that depends on distortions of 
the director field  with respect to
the uniform configuration aligned along the bipolar axis.

In general, the value of $Q_{\dd}$ varies depending
on a number of factors such as droplet size and shape,
anchoring conditions, applied voltage and so on.
In Ref.~\cite{Cox:1998}, such variations are found to be negligible
and the bipolar order parameter was estimated to be about $0.82$.
By contrast, the results of 
Refs.~\cite{Vic:mclc:1995,Vic:opt:1996}
suggest that the bipolar order parameter of a PDLC film
can be considerably affected by an applied electric voltage. 

In any event, so long as size and shape polydispersity
is weak, variations of the bipolar order parameter
throughout a sample can be reasonably disregarded.
From the other hand,
it is now commonly accepted that 
disorder in positions and orientation of bipolar droplets
is of vital importance for an understanding of
light scattering in PDLC 
systems~\cite{Zum:josa:1989,Kel:mclc:1994,Cox:1998}.
When the orientational distribution
of the bipolar axis is uniaxial, 
the relation of the form similar to Eq.~\eqref{eq:ord-drp}
\begin{equation}
  \label{eq:ord-ens}
  \avr{(3 \uvc{n}_{\dd}\otimes\uvc{n}_{\dd}-\vc{I})/2}_{d}
=Q(3 \uvc{N}\otimes\uvc{N}-\vc{I})/2
\end{equation}
describes this distribution in terms of 
the order parameter $Q$ and the optical axis $\uvc{N}$.

When direction of the bipolar axis is randomly distributed,
the PDLC film is isotropic and $Q=0$.  
This case is illustrated in Fig.~\ref{fig:disord}.
The initial off state
of a PDLC film is often assumed to be isotropic.

Applying an external electric field
will reorient NLC inside the droplets so as to introduce 
an overall anisotropy of the sample
with the optical axis $\uvc{N}$ parallel to the field.
Fig.~\ref{fig:ord} schematically shows a distribution of this sort.
The limiting case where all the droplets are aligned along the field
direction corresponds to the on state of a PDLC film
in the regime of saturation
(the saturated state) with $Q=1$.

The effect of the order parameters $Q_{\dd}$ and $Q$ on light scattering
in PDLC films will be of our major concern.

\subsection{Effective dielectric tensor}
\label{subsec:effect-diel-tens}

The films under consideration 
exemplify an inhomogeneous medium
in which a host material and isolated inclusions are
clearly identified. 
Our first step is to determine effective dielectric
characteristics of the film in the long wavelength limit.
To this end, we restrict ourselves to 
the case where the droplets are well
separated and the volume fraction of droplets, $\eta$,
is not too high. 

Under these conditions, following Refs.~\cite{Cox:1996,Levy:2000}, 
we may use the excluded volume approximation
discussed by Landauer in Ref.~\cite{Land:1978}
and write the total average electric field $\vc{E}$
as
\begin{equation}
  \label{eq:tot-avr-E}
  \vc{E}=\eta\vc{E}_{\dd}+(1-\eta)\vc{E}_p,
\end{equation}
where 
%% $E=U/d$, $U$ is the voltage across the film,
%% $d$ is the thickness of the film,
$\vc{E}_p$ and $\vc{E}_{\dd}$ 
are the average electric fields in the polymer matrix
and inside the droplets, respectively.
Using the relation for the mean electric polarization vectors:
$\vc{P}$, $\vc{P}_{\dd}$ and $\vc{P}_p$, that
can be written by analogy with Eq.~\eqref{eq:tot-avr-E}, we have
\begin{equation}
  \label{eq:tot-avr-P}
    \bs{\epsilon}_{\eff}\vc{E}=
\eta\bs{\epsilon}_{\dd}\vc{E}_{\dd}+(1-\eta)\epsilon_p\vc{E}_p.
\end{equation}
In Eq.~\eqref{eq:tot-avr-P} 
the orientationally averaged dielectric tensor~\eqref{eq:eps-drp}
is used as an approximation for
the dielectric tensor averaged over the droplets.

The fields $\vc{E}_{\dd}$ and $\vc{E}_p$
can now be related by means of an anisotropic version of the
traditional Maxwell-Garnett closure~\cite{Cox:1996,Levy:2000}
(reviews of other approximate schemes
can be found in~\cite{Beek:1967,Land:1978,Berg:sol:1992}):
\begin{equation}
  \label{eq:cl-MG}
\vc{E}_p = \vc{M}\vc{E}_{\dd},\quad
\vc{M}=(2\epsilon_p\,\vc{I}+
\bs{\epsilon}_{\dd})/(3\epsilon_p).
\end{equation}

Combining Eqs.~\eqref{eq:tot-avr-E},~\eqref{eq:tot-avr-P}
and~\eqref{eq:cl-MG}
gives the final result for $\bs{\epsilon}_{\eff}$:
\begin{align}
  \label{eq:eps-eff1}
&
  \bs{\epsilon}_{\eff}=
\epsilon_p\vc{I}+
\eta\,\avr{(\bs{\epsilon}_{\dd}-\epsilon_p\vc{I})\vc{M}^{-1}}
\notag\\
&
\times
\bigl[
\eta\,\avr{\vc{M}^{-1}}+(1-\eta)\vc{I}
\bigr]^{-1}=
\epsilon_m\vc{I}+\Delta\epsilon_m\uvc{N}\otimes\uvc{N},
\end{align}
\begin{equation}
  \label{eq:eps-m}
  \epsilon_m=\epsilon_p
\frac{\epsilon_{\perp}+2\epsilon_p+2\eta\epsilon_1}{%
\epsilon_{\perp}+2\epsilon_p-\eta\epsilon_1}
\approx \epsilon_p+\eta\epsilon_1,
\end{equation}
\begin{equation}
  \label{eq:Deps-m}
  \Delta\epsilon_m=9\epsilon_p^2\epsilon_a\eta\,
\frac{\epsilon_{\perp}+2\epsilon_p}{%
\epsilon_{\parallel}+2\epsilon_p}\,
\Bigl[
(\epsilon_{\perp}+2\epsilon_p-\eta\epsilon_1)
(\epsilon_{\perp}+2\epsilon_p-\eta\epsilon_2)
\Bigr]^{-1},
\end{equation}
where
$
\epsilon_1=
\epsilon_{\perp}-\epsilon_p+
\epsilon_p\epsilon_a(1-Q)/(\epsilon_{\parallel}+2\epsilon_p)
$
and
$
\epsilon_2=
\epsilon_{\perp}-\epsilon_p+
\epsilon_p\epsilon_a(1+2Q)/(\epsilon_{\parallel}+2\epsilon_p)
$. 

For the polymer with $n_p=1.54$
and the liquid crystal mixture E7
with $n_o^{(lc)}=1.5216$ and
$n_e^{(lc)}=1.74$,
it can be verified that the linear relation
on the right hand side of Eq.~\eqref{eq:eps-m}
gives a good approximation for the volume fraction
dependence of the effective dielectric constant $\epsilon_m$.
This is a  modified version of
the mixing rule
$\epsilon_m=(1-\eta)\epsilon_p+\eta\epsilon_{\perp}$
used in Ref.~\cite{Cox:1998}.

Eq.~\eqref{eq:Deps-m} clearly indicates that, in general,
the anisotropic part of the effective dielectric tensor
does not vanish. 
But the effective anisotropy parameter 
$\Delta\epsilon_m/\epsilon_m$ is found to be well under $0.05$, so that 
the effective medium is weakly
anisotropic as compared to NLC inside the droplets,
where  
$\epsilon_a^{(lc)}/\epsilon_{\perp}^{(lc)}>0.3$.
For this reason, as in 
Refs.~\cite{Kel:mclc:1994,Vic:pre:1993,Vic:mclc:1995,Vic:opt:1996,Cox:1998},
anisotropy of the effective medium can be disregarded.

In our subsequent treatment of the light scattering problem
we shall use  the relation~\eqref{eq:eps-m} to 
account for the dependent scattering effects by
replacing the polymer matrix surrounding the droplets with
the effective medium.
It is, however, should be noted that 
the result~\eqref{eq:eps-m} is strictly valid only in the
Rayleigh limit where the scatterer size is much smaller than
the wavelength of light (an extended discussion can be found in 
Chap.~9 of Ref.~\cite{Mish}).
This result can also be derived from the self-consistency 
condition that requires vanishing 
the average scattering amplitude in the forward
direction for droplets embedded in the effective medium~\cite{Stro:1978}.
So, the effective dielectric constant~\eqref{eq:eps-m}
serves as the lowest order approximation of a 
coherent potential approach~\cite{Souk:1994a}. 

\begin{figure*}[!tbh]
%\vskip5mm
\centering
\resizebox{150mm}{!}{\includegraphics*{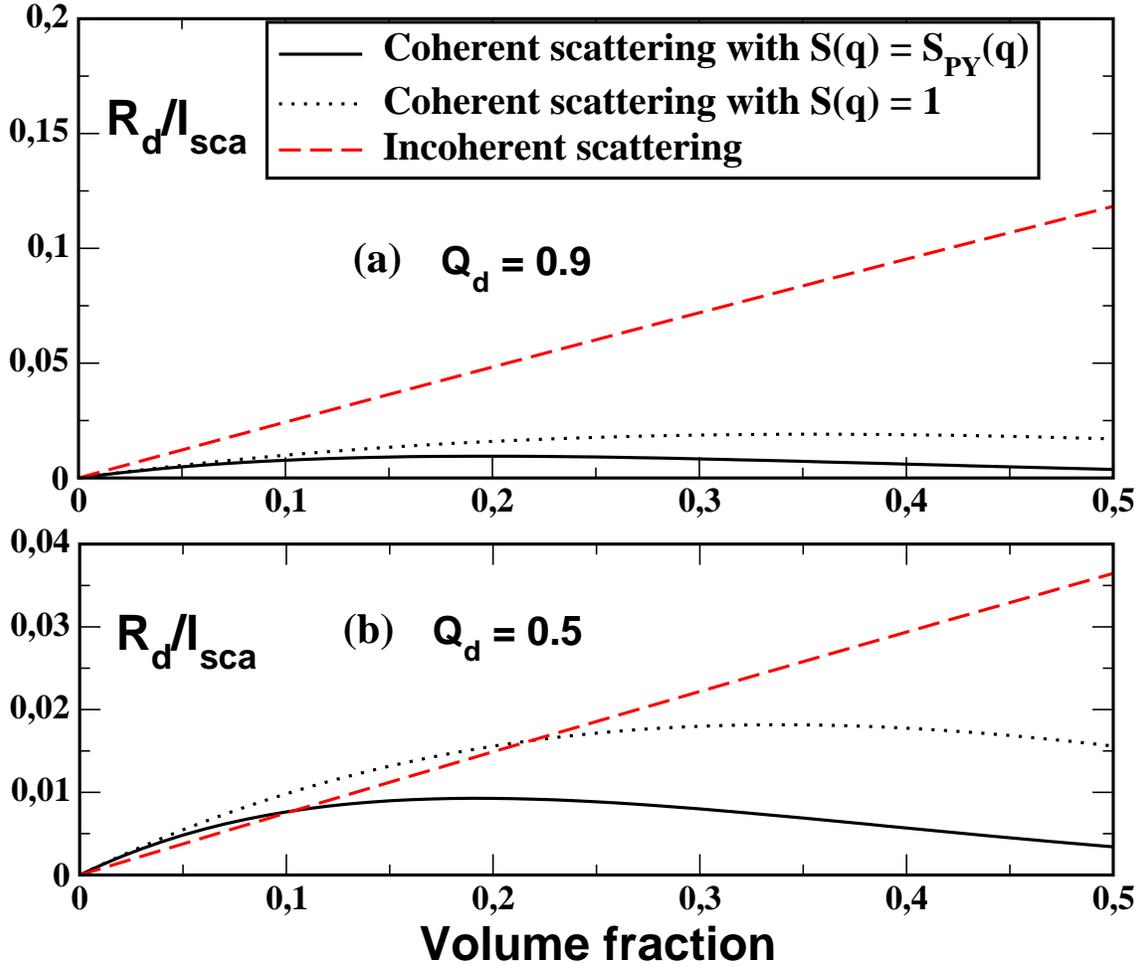}}
\caption{%
Coherent and incoherent light
scattering contributions to the ratio  $R_{\dd}/l_{sca}$ versus
the volume fraction $\eta$ at $R_{\dd}=0.5\mu$m 
for two values of 
the bipolar order parameter $Q_{\dd}$:
(a)~$Q_{\dd}=0.9$ and (b)~$Q_{\dd}=0.5$.
Refractive indices used in calculations are:
$n_p = 1.54$,  
$n_o^{(lc)}=1.52$ and
$n_e^{(lc)}=1.74$.    
}
\label{fig:sca-vf}
\end{figure*}

\begin{figure*}[!tbh]
%\vskip5mm
\centering
\resizebox{150mm}{!}{\includegraphics*{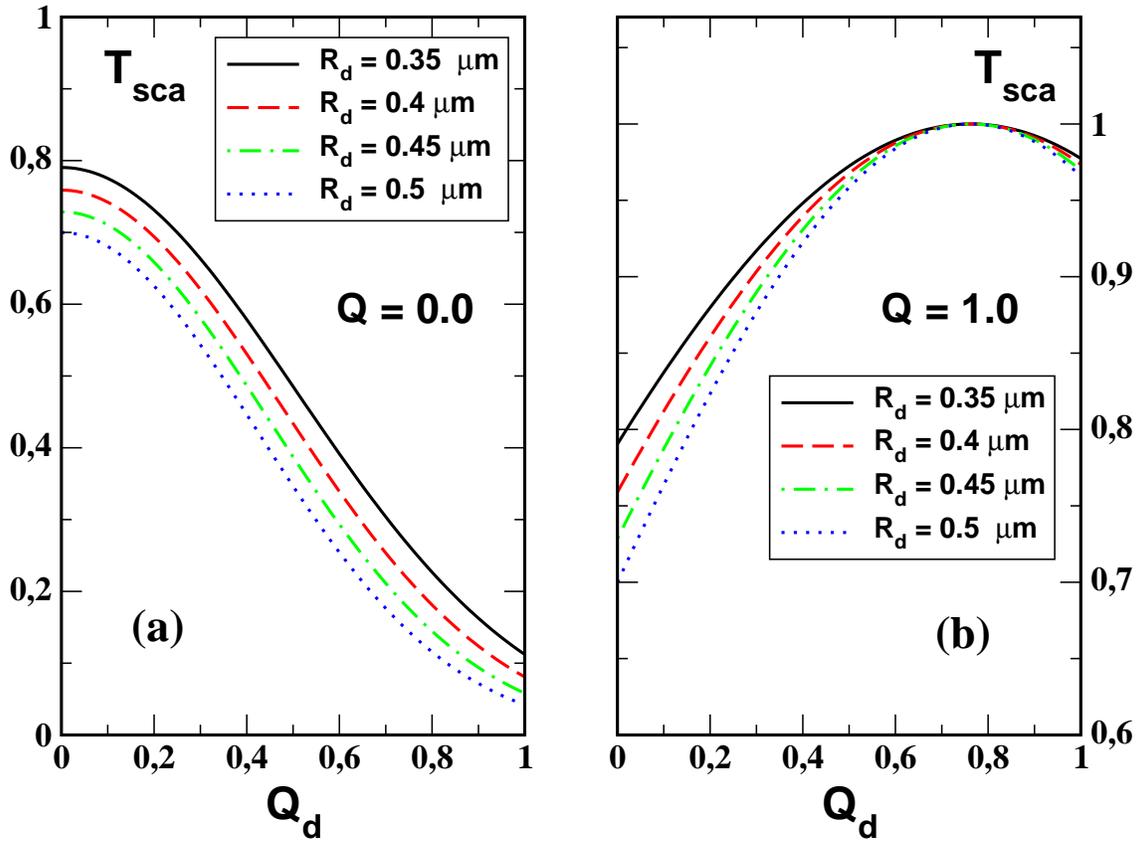}}
\caption{%
Transmittance as a function of the bipolar order parameter
$Q_{\dd}$ at the volume fraction $\eta=0.235$ for 
different droplet radii 
and two values of 
the order parameter $Q$:
(a)~$Q=0$ and (b)~$Q=1$.  
}
\label{fig:trans-Qd}
\end{figure*}

\begin{figure*}[!tbh]
%\vskip5mm
\centering
\resizebox{150mm}{!}{\includegraphics*{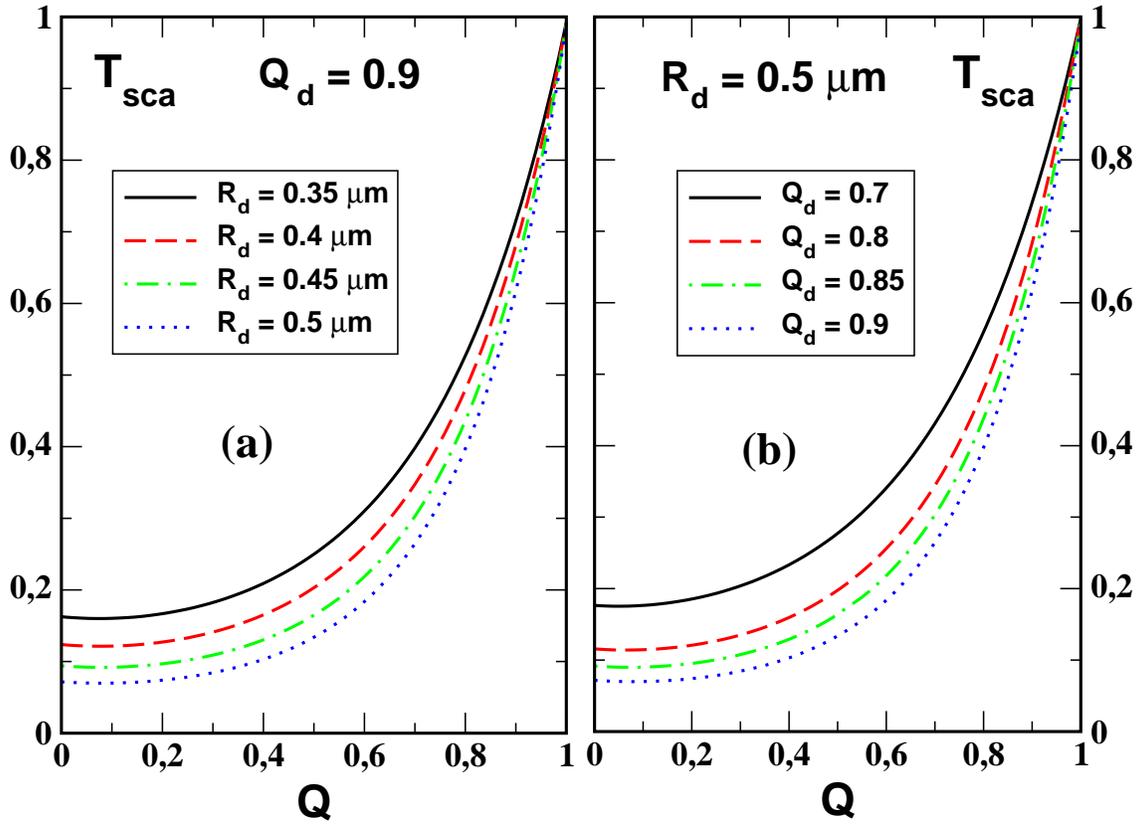}}
\caption{%
Transmittance as a function of the order parameter
$Q$ at 
(a)~various values of the mean droplet radius $R_{\dd}$
for $Q_{\dd}=0.9$ and
at (b)~various values of the bipolar order parameter $Q_{\dd}$
for $R_{\dd}=0.5\mu$m. The droplet volume fraction is $0.235$
and the thickness of PDLC layer is $20\,\mu$m.
}
\label{fig:trans-Q}
\end{figure*}

\subsection{Light scattering and optical transmittance}
\label{subsec:light-scatt}
  
In this section we begin with the light scattering problem for
a single LC droplet 
characterized by the average dielectric tensor~\eqref{eq:eps-drp}
and surrounded by the medium
with the effective refractive index $n_m=\sqrt{\epsilon_m}$.
Using the well-known Rayleigh-Gans approximation~\cite{Ishim,New},
we compute the elements of the scattering amplitude matrix
and the scattering cross section.

Then we apply the Percus-Yevick approximation~\cite{Perc:pr:1958,Zim:bk:1979}
to evaluate the effective scattering as 
the scattering cross section averaged
over positions of the droplets and perform averaging
of the scattering cross section
over the bipolar axis orientation.
Finally, in the low concentration approximation,
we derive the expressions for
the scattering mean free path and the optical transmittance.

\subsubsection{Scattering amplitude matrix in Rayleigh-Gans approximation}
\label{subsubsec:sca-ampl}
  
Following the standard procedure~\cite{Ishim,New},
the undisturbed incident wave is
assumed to be a harmonic plane wave 
with the frequency $\omega$ and the wavenumber
$k=n_m \omega/c$. 
The wave is propagating  through the effective medium
along the direction specified by a unit vector
$\uvc{k}_{inc}$ and
the wave vector $\vc{k}_{inc}=k\uvc{k}_{inc}$.
The polarization vector of the electric field is
\begin{equation}
  \label{eq:wav-inc}
\vc{E}_{inc}=
{E}_i^{(inc)}\,\uvc{e}_i(\uvc{k}_{inc}),
\end{equation}
where the basis vectors
$\uvc{e}_x(\uvc{k}_{inc})=
(\cos\theta_i\cos\phi_i,\cos\theta_i\sin\phi_i,-\sin\theta_i)$
and
$\uvc{e}_y(\uvc{k}_{inc})=
(-\sin\phi_i,\cos\phi_i,0)$ are perpendicular to $\uvc{k}_{inc}$
defined by  the polar and azimuthal angles:
$\theta_i$ and $\phi_i$.
[Throughout the paper summation over repeated indices will be assumed.]

Asymptotic behavior of the scattered outgoing wave
in the far field region ($kr\gg 1$)  
is known~\cite{Ishim,New}:
\begin{equation}
  \label{eq:wav-sca}
  \vc{E}_{sca}(\vc{r})\sim\vc{E}_{sca}
\frac{\e^{ikr}}{kr},\quad
\vc{E}_{sca}=
{E}_i^{(sca)}\,\uvc{e}_i(\uvc{k}_{sca}),
\end{equation}
where $\uvc{k}_{sca}=\uvc{r}=
(\sin\theta\cos\phi,\sin\theta\sin\phi,\cos\theta)$ and
$\vc{E}_{sca}$ is linearly related to 
the polarization vector of the incident
wave~\eqref{eq:wav-inc} through
the scattering amplitude matrix
$\vc{A}(\vc{k}_{sca},\vc{k}_{inc})$ in the following way:
\begin{equation}
  \label{eq:sca-ampl}
  {E}_i^{(sca)}=
A_{ij}(\vc{k}_{sca},\vc{k}_{inc}){E}_j^{(inc)}.
\end{equation}

The elements of the scattering amplitude matrix can be easily
computed by using the Rayleigh-Gans
approximation (RGA)~\cite{Ishim,New}.
This approximation is known to be applicable 
to the case of submicron nematic droplets~\cite{Zum:1986,Cox:1998,Lecl:1999}
and gives the following result:
\begin{equation}
  \label{eq:ampl-RGA-gen}
 A_{ij}(\uvc{k}_{sca},\uvc{k}_{inc})
=
\frac{V_{\dd} k^3}{4\pi}\, 
\uvc{e}_i(\uvc{k}_{sca})\cdot
\bs{\epsilon}(\vc{q})\cdot
\uvc{e}_j(\uvc{k}_{inc}),
\end{equation}
\begin{equation}
  \label{eq:eps-q}
  \bs{\epsilon}(\vc{q})=
V_{\dd}^{-1}\int_{V_{\dd}}\bs{\epsilon}\,\e^{i(\vc{q}\cdot\vc{r})}\dd^3\vc{r}=
f(qR_{\dd})\,\bs{\epsilon},
\end{equation}
\begin{equation}
  \label{eq:eps-dif}
  \bs{\epsilon}=(\bs{\epsilon}_{\dd}-\epsilon_m\vc{I})/\epsilon_m\equiv
\alpha\vc{I}+\beta\uvc{n}_{\dd}\otimes\uvc{n}_{\dd},
\end{equation}
where $V_{\dd}=4\pi R_{\dd}^3/3$ is the droplet volume; 
$R_{\dd}$ is the droplet radius,
$\vc{q}=\vc{k}_{sca}-\vc{k}_{inc}$,
$f(x)=3(\sin x- x\cos x)/x^3$,
$\alpha\equiv \epsilon_{\perp}/\epsilon_m-1$,
$\beta\equiv \epsilon_{a}/\epsilon_m$.

\subsubsection{Scattering cross section}
\label{subsubsec:cross-sec}

All scattering properties of a droplet can be computed from
the elements of the scattering amplitude
matrix~\eqref{eq:ampl-RGA-gen}.
In particular, when the incident wave is unpolarized,
it is not difficult to deduce the expression for the differential
scattering cross section characterizing the angular distribution of
the scattered light:
\begin{equation}
  \label{eq:cross-sec-diff}
  \sigma_{\mrm{diff}}=
\frac{A_{\dd}(kR_{\dd})^4}{18\pi} F(qR_{\dd})\,\psi_{ij}\psi_{ij},\quad
\psi_{ij}=\uvc{e}_i(\uvc{k}_{sca})\cdot
\bs{\epsilon}\cdot
\uvc{e}_j(\uvc{k}_{inc}),
\end{equation}
where 
$F(qR_{\dd})=f^2(qR_{\dd})$ is the form factor and
$A_{\dd}=\pi R_{\dd}^2$ is the area of the droplet projection
onto the plane normal to $\uvc{k}_{inc}$.

The RGA scattering cross section~\eqref{eq:cross-sec-diff} depends
on droplet orientation only through the last factor
which can be rewritten in the following form:
\begin{equation}
  \label{eq:psi-gen}
  \psi_{ij}\psi_{ij}\equiv\psi^2=
\Tr\bs{\epsilon^2}+
[\uvc{k}_{sca}\,\bs{\epsilon}\,\uvc{k}_{inc}]^2-
\uvc{k}_{inc}\,\bs{\epsilon}^2\,\uvc{k}_{inc}-
\uvc{k}_{sca}\,\bs{\epsilon}^2\,\uvc{k}_{sca}.
\end{equation}
Substituting the tensor~\eqref{eq:eps-dif}
into Eq.~\eqref{eq:psi-gen} yields 
the orientationally dependent factor in
the explicit form:
\begin{align}
  \label{eq:psi-gen2}
&
  \psi^2=\alpha^2 
\bigl[
1+\sca{\uvc{k}_{inc}}{\uvc{k}_{sca}}^2
\bigr]+
\bigl[
(\alpha+\beta)^2-\alpha^2
\bigr]
\notag\\
&
\times
\bigl[
1-\sca{\uvc{k}_{inc}}{\uvc{n}_{\dd}}^2
-\sca{\uvc{k}_{sca}}{\uvc{n}_{\dd}}^2
\bigr] + 
\bigl[
\beta
\sca{\uvc{k}_{inc}}{\uvc{n}_{\dd}}
\sca{\uvc{k}_{sca}}{\uvc{n}_{\dd}}
\bigr]^2
\notag
\\
&
+ 2\alpha\beta
\sca{\uvc{k}_{inc}}{\uvc{k}_{sca}}
\sca{\uvc{k}_{inc}}{\uvc{n}_{\dd}}
\sca{\uvc{k}_{sca}}{\uvc{n}_{\dd}}.
\end{align}
This result will be used to perform orientational averaging
of the scattering cross section.

\subsubsection{Scattering mean free path}
\label{subsubsec:mean-free}

Our task now is to evaluate the scattering mean free path,
$l_{sca}$, which 
is also known as the phase coherence length 
or the extinction length~\cite{Souk:1994a,Souk:1994b}.
This is the characteristic distance between two scattering events 
after which the phase coherence of radiation
gets lost leading to the exponential decay of the incident intensity
known as the Lambert-Beer law~\cite{Shen:1995,Ross:1999}.
So, the optical transmittance through a film of thickness $d$ is
given by
\begin{equation}
  \label{eq:transmit}
  T_{sca}=\exp(-d/l_{sca}).
\end{equation}

So long as the volume fraction is not too high,
the scattering mean free path can be derived by using 
the low concentration approximation. The result is
\begin{equation}
  \label{eq:lsca}
  l_{sca}^{-1}=n\int\sigma_{\eff}\,\dd\uvc{k}_{sca},
\end{equation}
where
$\dd\uvc{k}_{sca}\equiv \sin\theta\dd\theta\dd\phi$ and
$n=\eta/V_{\dd}$ is the number density of the droplets.
In the limit of weak size and shape polydispersity, we define
the effective cross section $\sigma_{\eff}$
as the single scattering cross section~\eqref{eq:cross-sec-diff}
averaged over both positions and orientation of bipolar droplets:
\begin{equation}
  \label{eq:cross-sec-eff}
\sigma_{\eff}=
  \avr{\avr{\sigma_{\mrm{diff}}}}=
\frac{A_{\dd}(kR_{\dd})^4}{18\pi} F(qR_{\dd})\,\avr{\avr{\psi^2}},
\end{equation}
where $R_{\dd}$ is now the mean value of the droplet radius.
In order to avoid ambiguity, we shall denote averages over droplet
type by $\avr{\avr{\ldots}}$ and averages 
over bipolar axis orientation by $\avr{\dots}$.

When positional and orientational degrees of freedom are
statistically independent, the result
of averaging over positional disorder
is known to be a sum of two terms 
describing the coherent and the incoherent
scattering~\cite{Lax:1951,Zim:bk:1979,Cox:1998}. So, we have
\begin{equation}
  \label{eq:psi-decomp}
  \avr{\avr{\psi^2}}\equiv\Psi=\Psi_{\mrm{coh}}+\Psi_{\mrm{incoh}},\quad
\Psi_{\mrm{coh}}=S(\vc{q})\,\avr{\psi}^2,\quad
\Psi_{\mrm{incoh}}=\avr{\psi^2}-\avr{\psi}^2,
\end{equation}
where the contributions to the coherent and the incoherent scattering
are proportional to 
$\Psi_{\mrm{coh}}$ and $\Psi_{\mrm{incoh}}$, respectively;
$S(\vc{q})=1+n\int\e^{i(\vc{q}\cdot\vc{r})} 
[g(\vc{r})-1]\,\dd^3\vc{r}$ is the structure factor
expressed in terms of the pair correlation function 
$g(\vc{r})$~\cite{Zim:bk:1979}.
   
The coherent part of $\Psi$ describes light scattering by a
positionally disordered ensemble of identical droplets. The
scattering properties of each droplet are now characterized by
the orientationally averaged tensor~\eqref{eq:eps-dif}:
\begin{equation}
  \label{eq:avr-eps-dif}
    \avr{\bs{\epsilon}}=
\bar{\alpha}\vc{I}+\bar{\beta}\uvc{N}\otimes\uvc{N}
=\bigl[
\alpha+\beta (1-Q)/3\bigr]
\vc{I}+\beta Q \uvc{N}\otimes\uvc{N}
\end{equation}
and we obtain $\avr{\psi}^2$ from the expression~\eqref{eq:psi-gen2}
modified as follows 
\begin{align}
  \label{eq:avr-psi2}
&
  \avr{\psi}^2=\bar{\alpha}^2 
\bigl[
1+\sca{\uvc{k}_{inc}}{\uvc{k}_{sca}}^2
\bigr]+
\bigl[
(\bar{\alpha}+\bar{\beta})^2-\bar{\alpha}^2
\bigr]
\notag\\
&
\times
\bigl[
1-\sca{\uvc{k}_{inc}}{\uvc{N}}^2
-\sca{\uvc{k}_{sca}}{\uvc{N}}^2
\bigr] + 
\bigl[
\bar{\beta}
\sca{\uvc{k}_{inc}}{\uvc{N}}
\sca{\uvc{k}_{sca}}{\uvc{N}}
\bigr]^2
\notag
\\
&
+ 2\bar{\alpha}\bar{\beta}
\sca{\uvc{k}_{inc}}{\uvc{k}_{sca}}
\sca{\uvc{k}_{inc}}{\uvc{N}}
\sca{\uvc{k}_{sca}}{\uvc{N}}.
\end{align}

When the structure factor equals unity, $S(\vc{q})=1$,
the droplets are positionally uncorrelated.
This approximation, however, cannot be appropriate 
for the morphology, where the droplets do not overlap.
As in Ref.~\cite{Cox:1998},
in order to take into account droplet self-avoidance,  
we shall use the structure factor for binary mixtures of hard spheres
computed from the exact solution of 
the Percus-Yevick equation~\cite{Perc:pr:1958,Zim:bk:1979}.
The structure factor expressed in terms of the Ornstein-Zernike
direct correlation function $c(r)$ 
is given by~\cite{Thiele:jcp:1963,Werth:prl:1963,Lebow:pra:1964} 
\begin{align}
  \label{eq:PY1}
&
  S(\vc{q})=S_{\mrm{PY}}(qR_{\dd})=\frac{1}{1-n C(\vc{q})},
\quad
C(\vc{q})=
\int\e^{i(\vc{q}\cdot\vc{r})} c(r)\,\dd^3\vc{r},
\\
\label{eq:dir-corr}
&
c(r)=
\begin{cases}
6\eta\gamma_1 (r/2R_{\dd})-\gamma_2[1+\eta/2(r/2R_{\dd})^3], & r\le 2R_{\dd},\\
0, & r>2R_{\dd},
\end{cases}
\end{align}
where 
$\gamma_1=(1+\eta/2)^2/(1-\eta)^4$
and
$\gamma_2=(1+2\eta)^2/(1-\eta)^4$.
%% Rather straightforward algebra yields 
%% the explicit formula used in our calculations
%% \begin{align}
%% \label{eq:s-fac-PY}
%% &
%% S_{\mrm{PY}}(x)=\frac{(1-\eta)^4}{%
%% (1-\eta)^4+\eta\bigl[ (1+2\eta)^2(8 f_1(2x)+4\eta f_4(2x))
%% -9\eta (2+\eta)^2 f_2(2x)
%% \bigr]},
%% \end{align}
%% where $f_n(x)=(n+2)\, x^{-n-2}\int_{0}^{x} t^n\sin t\,\dd t$.

Droplet orientation fluctuations induce
the incoherent scattering described by $\Psi_{\mrm{incoh}}$ in
Eq.~\eqref{eq:psi-decomp}.
The expression for $\Psi_{\mrm{incoh}}$ can be derived in the following form
\begin{align}
  \label{eq:psi-incoh}
&
  \Psi_{\mrm{incoh}}=\frac{\beta^2}{9}\,(1-Q)
\bigl[ 3(1+Q)-(1-Q)
\sca{\uvc{k}_{inc}}{\uvc{k}_{sca}}^2
\bigr]
\notag\\
&
+\beta^2
\Bigl[
\avr{\sca{\uvc{k}_{inc}}{\uvc{n}_{\dd}}^2
\sca{\uvc{k}_{sca}}{\uvc{n}_{\dd}}^2}
-
\avr{\sca{\uvc{k}_{inc}}{\uvc{n}_{\dd}}^2}
\avr{\sca{\uvc{k}_{sca}}{\uvc{n}_{\dd}}^2}
\Bigr],
\end{align}
where
$
\avr{\sca{\uvc{k}_{inc (sca)}}{\uvc{n}_{\dd}}^2}=
Q\sca{\uvc{k}_{inc (sca)}}{\uvc{N}}^2+(1-Q)/3
$.
In accordance with the above interpretation, Eq.~\eqref{eq:psi-incoh}
shows that the incoherent scattering is solely caused by
the anisotropic part of the droplet dielectric tensor
and disappears when the droplets are perfectly aligned  with $Q=1$.
In addition to the order parameter $Q$, 
there are fourth order averages in 
the last square bracketed term on the right hand side
of Eq.~\eqref{eq:psi-incoh} and we need to know
additional higher order parameters to estimate 
the incoherent scattering.

The special case in which
both the incidence direction and the optical axis of the film
are normal the substrates, 
$\uvc{k}_{inc}=\uvc{N}=\uvc{e}_z$, is  
of our particular interest.
In this case 
the above results take the simplified form:
\begin{equation}
  \label{eq:psi-coh-norm}
  \Psi_{\mrm{coh}}(\theta)=\bar{\alpha}^2\,S(\vc{q})\,
\bigl[
1+\cos^2\theta
\bigr],
\end{equation}
\begin{align}
  \label{eq:psi-incoh-norm}
&
  \Psi_{\mrm{incoh}}(\theta)=\frac{\beta^2}{9}\,(1-Q)
\bigl[ 3(1+Q)-(1-Q)\cos^2\theta
\bigr]
\notag\\
&
+\beta^2\,\frac{3\cos^2\theta-1}{2}\,
\Bigl[
\avr{n_z^4}
-
\avr{n_z^2}^2
\Bigr],
\end{align}
where $\cos\theta=\sca{\uvc{k}_{sca}}{\uvc{e}_z}$
and
$n_z=\sca{\uvc{n}_{\dd}}{\uvc{e}_z}$.

We can now combine
Eqs.~\eqref{eq:lsca}-\eqref{eq:psi-decomp}
and the relations~\eqref{eq:psi-coh-norm}-\eqref{eq:psi-incoh-norm}
to obtain 
the ratio of the droplet radius $R_{\dd}$ and the scattering
mean free path $l_{sca}$:
\begin{equation}
  \label{eq:scat-path}
  R_{\dd}/l_{sca}=\frac{\eta (kR_{\dd})^4}{12}
\int_{-1}^{1} F(qR_{\dd})\,\Psi(\theta)\,\dd(\cos\theta),
\end{equation}
where $q^2=2 k^2 (1-\cos\theta)$.

%% In addition to the scattering mean free path,
%% we may also consider
%% the transport
%% mean free path $l_{tr}$ defined as the length over which
%% momentum transfer becomes uncorrelated~\cite{Souk:1994a,Souk:1994b}.
%% This is the mean distance after which the direction of radiation
%% gets lost and which plays an important role in the theory of diffusive wave
%% transport~\cite{Shen:1995,Ross:1999}.
%% The ratio of $R_{\dd}$ and $l_{tr}$ is 
%% \begin{equation}
%%   \label{eq:transp-path}
%%   R_{\dd}/l_{tr}=\frac{\eta (kR_{\dd})^4}{12}
%% \int_{-1}^{1} F(qR_{\dd})\,\Psi(\theta)\,(1-\cos\theta)\,\dd(\cos\theta)
%% \end{equation}
%% and it might be expected that, 
%% similarly to  Eq.~\eqref{eq:transmit}, the relation
%% \begin{equation}
%%   \label{eq:trans-tr}
%%   T_{tr}=\exp(-d/l_{tr})
%% \end{equation}
%% estimates the optical transmittance through a film 
%% in the diffusive regime~\cite{Cox:1998}.

\subsection{Numerical results}
\label{subsec:num-res}

Optical transmittance of a normally incident unpolarized
light through a film can be computed 
from Eq.~\eqref{eq:transmit} combined  
with the formulae~\eqref{eq:PY1}
and~\eqref{eq:psi-coh-norm}-\eqref{eq:scat-path}.

Fig.~\ref{fig:sca-vf} shows that correlations in droplet
positions generally reduce the incoherent scattering
as compared to the case of positionally uncorrelated droplets.
This effect becomes more pronounced as the bipolar order parameter 
decreases. Referring to Fig.~\ref{fig:sca-vf}(a), the incoherent
scattering dominates at relatively large bipolar order parameters
and, as is seen from Fig.~\ref{fig:sca-vf}(b), this is no longer the
case for weakly anisotropic droplets.

Dependencies of the optical transmittance on
the bipolar order parameter for different droplet radii
are given in Fig.~\ref{fig:trans-Qd}.
When the angular distribution of the bipolar axis is 
isotropic and $Q=0$, Fig.~\ref{fig:trans-Qd}(a) shows
that the transmittance declines as $Q_{\dd}$ increases.
Such behaviour is mainly due to
an increase in the incoherent scattering
governed by the anisotropic part of the tensor~\eqref{eq:eps-dif}.
The latter is the ratio $\beta=\epsilon_a/\epsilon_m$ that 
characterizes anisotropy of the bipolar droplets
and grows with the bipolar order parameter.

In the opposite case of
perfectly aligned droplets with $Q=1$, shown in
Fig.~\ref{fig:trans-Qd}(b), the incoherent scattering is suppressed.
It is seen that the transmittance reaches 
its maximal value at $Q_{\dd}\approx 0.8$
where  the refractive index of the effective medium
$n_m$ and the ordinary refractive index of the droplets 
$n_o=\sqrt{\epsilon_{\perp}}$ are matched.
Equivalently, from Eq.~\eqref{eq:psi-coh-norm} 
the coherent scattering disappears
when the matching condition: $\bar{\alpha}=0$ is fulfilled.

Computing the dependence of the transmittance on the order parameter $Q$
requires the knowledge of the variance
$\avr{n_z^4}-\avr{n_z^2}^2$ that enters the factor~\eqref{eq:psi-incoh-norm} 
describing the incoherent scattering.
The variance of $n_z^2$ cannot be negative and
vanishes at both $Q=1$ and $Q=-1/2$.  In addition, 
it  equals $4/45$ in the isotropic state with $Q=0$.  

For simplicity, we suppose that
the orientational distribution can be parameterized by the order
parameter $Q$ and approximate the variance by
the simple polynomial 
with a maximum at $Q=0$:
$\avr{n_z^4}-\avr{n_z^2}^2\approx 4(1-Q)^2 (2Q+1)/45$.
The results of numerical calculations are presented   
in Fig.~\ref{fig:trans-Q}.

\section{Discussion}
\label{sec:discussion}

Dependencies of the transmittance on the order
parameter $Q$ plotted in Fig.~\ref{fig:trans-Q}(b)
can be used to estimate the values of $Q$ and $Q_{\dd}$
from the transmission coefficient measured 
in the zero-field and saturated states,
$T_0$ and $T_{sat}$,
of PDLC films prepared in the presence of an electrical field. 
The first column of Table~\ref{tab:T-Q-vs-U}
gives the voltage $U_{uv}$ applied 
across the film of PDLC precursor during UV polymerization.
This voltage affects ordering of the droplets 
and the experimental results for $T_0$ and $T_{sat}$ 
are listed in the second and third columns, respectively.

According to Table~\ref{tab:T-Q-vs-U}, the smallest value
of $T_0$, $T_0\approx 0.08$,
corresponds to the sample prepared in the absence
of external fields. 
Theoretically, this value can be estimated
as the minimal transmittance of the $T_{sca}$ vs $Q$ curve.
Referring to Fig.~\ref{fig:trans-Q}(b), this transmittance
depends on the bipolar order parameter $Q_{\dd}$.
The theoretical results are found to be in agreement with
the experimental value $T_0\approx 0.08$
at $Q_{\dd}$ ranged between $0.85$ and $0.9$.
So, the bipolar order parameter appears to be close to
the estimate of Ref.~\cite{Cox:1998}, $Q_{\dd}\approx 0.82$.

We also found that 
the order parameter $Q$ corresponding to the
transmittance $T_0$ at $U_{uv}=0$ is very small ($Q\approx 0.04$).
This result supports the commonly accepted assumption 
that  PDLC films prepared with no
applied voltage are nearly isotropic. 

%% This assumption, however, has not been tested experimentally yet. 

When $U_{uv}\ne 0$ and $Q_{\dd}$ is known, 
the initial transmittance $T_0$ and the order parameter $Q$
can be evaluated from the corresponding $T_{sca}$ vs $Q$ curve.
For $Q_{\dd}=0.85$ and $Q_{\dd}=0.9$, 
the results are given in the last four columns of
Table~\ref{tab:T-Q-vs-U}.
The order parameter is shown to be an increasing function
of $U_{uv}$. When the voltage $U_{uv}$ increases, 
the value of $Q$ rapidly grows and saturates.

It should be emphasized that, in addition to the order parameters
$Q_{\dd}$ and $Q$,
there are a number of parameters that enter the theory.
For these parameters
we have used the values estimated from the experimental data.
These are: the refractive indices of 
the polymer ($n_p=1.54$) 
and the liquid crystal ($n_o^{(lc)}=1.52$ and $n_e^{(lc)}=1.74$),
the thickness of the film ($d=20\,\mu$m), 
the mean value of the droplet
radius ($R_{\dd}=0.5\,\mu$m) 
and the volume fraction of droplets ($\eta=0.235$).

Finally, we comment on limitations of our theoretical treatment
that essentially relies on the Rayleigh-Gans approximation
to describe light scattering in terms of the order parameters.
It is, however, applicable only for sufficiently small droplets
when $|n_{o,e}/n_m-1|\ll 1$ and
$kR_{\dd}|n_{o,e}/n_m-1|\ll 1$.

The single scattering by larger droplets
has been previously studied by using
the anomalous diffraction approach~\cite{Zum:1988}
and the \textit{T}-matrix theory~\cite{Kis:pre:2002}.
In these more complicated cases averaging over droplet
orientation will require the knowledge of additional higher order
averages characterizing orientational distribution of the droplets.
Nevertheless,
as it can be concluded from the experimental 
data of Sec.~\ref{sec:experiment},
the Rayleigh-Gans approximation is acceptable
provided the droplet radius is below $0.7\mu$m.

We have also used the effective medium theory combined with
the low concentration approximation. For optically
isotropic scatterers, such a mixed approach
is known to provide reasonably accurate results
when the volume fraction is under $0.3$~\cite{Souk:1994b}.
Otherwise, a more sophisticated treatment of multiple scattering
effects is required.

\section{Conclusion}
\label{sec:concl}

In this study we have considered 
light scattering in PDLC composites with the
``Swiss cheese'' morphology in relation to the orientational
order parameters of bipolar droplets.
We have applied the theoretical approach
based on the Rayleigh-Gans approximation
to estimate the order parameters from the 
experimental results on  light transmittance 
through the sample. 

It is shown that the droplet ordering can be controlled 
by applying an electrical field during phase
separation.
This ordering increases with the voltage and saturates. 
In the samples prepared in the absence of the field, 
the droplets are found to be almost randomly distributed.

%% The developed approach is not limited to the experimental conditions
%% used in the present studies. It can be also applied to estimate the
%% drop's order in the samples with in-plane orientation of LC drops
%% (most desirable case from the point of contrast
%% enhancement). Moreover, it can be applied to calculate angular
%% characteristics of the sample transmittance. ..

\begin{acknowledgments}
This study was carried out under the project
``Ordering regularities and properties of nano-composite systems''
supported by the National Academy of Sciences of Ukraine.
We also thank O.~Lavrentovich and Liou Qiu (Kent Liquid Crystal
Institute) for assistance with SEM measurements.
\end{acknowledgments}

%\bibliographystyle{apsrev}
%\bibliography{polymer,scatter,lc,quant,qft}

\end{document}